\documentclass[aps,prb,twocolumn,showpacs,superscriptaddress]{revtex4-1}

\usepackage{graphicx}

\begin{document}

\title{Experimental elucidation of the origin of the `double spin resonances' in Ba(Fe$_{1-x}$Co$_x$)$_2$As$_2$}
\author{Meng Wang}
\email{wangm@berkeley.edu}
\affiliation{Department of Physics, University of California, Berkeley, California 94720, USA }
\author{M. Yi}
\email{mingyi@berkeley.edu}
\affiliation{Department of Physics, University of California, Berkeley, California 94720, USA }
\author{H. L. Sun}
\affiliation{Beijing Synchrotron Radiation Facility, Institute of High Energy Physics, Chinese Academy of Science, Beijing 100049, China}
\author{P. Valdivia}
\affiliation{Department of Physics, University of California, Berkeley, California 94720, USA }
\author{M. G. Kim}
\affiliation{Materials Science Division, Lawrence Berkeley National Laboratory, Berkeley, California 94720, USA }
\author{Z. J. Xu}
\affiliation{Department of Physics, University of California, Berkeley, California 94720, USA }
\author{T. Berlijn}
\affiliation{Center for Nanophase Materials Sciences and Computer Science and Mathematics Division, Oak Ridge National Laboratory, Oak
Ridge, Tennessee 37831, USA }
\author{A. D. Christianson}
\affiliation{Quantum Condensed Matter Division, Oak Ridge National Laboratory, Oak
Ridge, Tennessee 37831, USA}
\author{Songxue Chi}
\affiliation{Quantum Condensed Matter Division, Oak Ridge National Laboratory, Oak
Ridge, Tennessee 37831, USA}
\author{M. Hashimoto}
\affiliation{Stanford Synchrotron Radiation Lightsource, SLAS National Accelerator Laboratory, Menlo Park, California 94025, USA}
\author{D. H. Lu}
\affiliation{Stanford Synchrotron Radiation Lightsource, SLAS National Accelerator Laboratory, Menlo Park, California 94025, USA}
\author{X. D. Li}
\affiliation{Beijing Synchrotron Radiation Facility, Institute of High Energy Physics, Chinese Academy of Science, Beijing 100049, China}
\author{E. Bourret-Courchesne}
\affiliation{Materials Science Division, Lawrence Berkeley National Laboratory, Berkeley, California 94720, USA }
\author{Pengcheng Dai}
\affiliation{Department of Physics and Astronomy, Rice University, Houston, Texas 77005, USA}
\author{D. H. Lee}
\affiliation{Department of Physics, University of California, Berkeley, California 94720, USA }
\affiliation{Materials Science Division, Lawrence Berkeley National Laboratory, Berkeley, California 94720, USA }
\author{T. A. Maier}
\affiliation{Center for Nanophase Materials Sciences and Computer Science and Mathematics Division, Oak Ridge National Laboratory, Oak
Ridge, Tennessee 37831, USA }
\author{R. J. Birgeneau}
\affiliation{Department of Physics, University of California, Berkeley, California 94720, USA }
\affiliation{Materials Science Division, Lawrence Berkeley National Laboratory, Berkeley, California 94720, USA }
\affiliation{Department of Materials Science and Engineering, University of California, Berkeley, California 94720, USA }

\begin{abstract}
We report a combined study of the spin resonances and superconducting gaps for underdoped ($T_c=19$ K), optimally doped ($T_c=25$ K), and overdoped ($T_c=19$ K) Ba(Fe$_{1-x}$Co$_x$)$_2$As$_2$ single crystals with inelastic neutron scattering and angle resolved photoemission spectroscopy. We find a quasi two dimensional spin resonance whose energy scales with the superconducting gap in all three compounds. In addition, anisotropic low energy spin excitation enhancements in the superconducting state have been deduced  and characterized for the under and optimally doped compounds. Our data suggest that the quasi two dimensional spin resonance is a spin exciton that corresponds to the spin singlet-triplet excitations of the itinerant electrons. However, the intensity enhancements of the anisotropic spin excitations are dominated by the out-of-plane spin excitations of the ordered moments due to the suppression of damping in the superconducting state. Hence we offer a new interpretation of the double energy scales differing from previous interpretations based on anisotropic superconducting energy gaps, and systematically explain the doping-dependent trend across the phase diagram.

\end{abstract}

\maketitle




\section{Introduction}

A spin resonance mode in an unconventional superconductor is a spin exciton that corresponds to the spin singlet-triplet excitation of itinerant electrons within the superconducting (SC) gap energy\cite{Eschrig2006,Dai2015,Inosov2016}. In an inelastic neutron scattering (INS) spectrum, one can see a dramatic enhancement in the intensity for spin excitations at a specific energy $\omega$ and momentum $Q$ below the SC transition temperature $T_c$. The energy and momentum of the spin resonance mode have intimate relationships with the magnitude and symmetry of the SC gap. Thus, the spin resonance mode has been viewed as crucial evidence for unconventional superconductivity and as a probe for revealing the SC gap symmetry. However, while the intensity enhancement of the spin excitations for a superconductor in the SC state can be ascribed to the spin exciton mode, it can also originate from the suppression of damping by superconductivity on a pre-existing magnon mode of an underlying antiferromagnetic (AF) order\cite{Pailhes2006}.

In the iron pnictide superconductors, the spin resonance mode has been observed at the coincident wave vectors of the nesting wave vector $(\pi, 0)$ between the hole and electron Fermi surfaces (FSs) and the long range AF order $Q_{AF}$ of the parent compounds with an energy $E_{res}\approx4.3k_BT_c$\cite{Christianson2008,Chi2009,Li2009,Lumsden2009,Yu2009,Park2010,Inosov2011,Terashima2009,Wang2010,Zhang2011,Liu2012a,Luo2012a}. This observation is consistent with the prediction for an $s\pm$ pairing symmetry where the spin resonance appears below the partical-hole spin flip continuum $2\Delta=|\Delta(k+Q_{AF})+\Delta(k)|\approx6k_BT_c$, where $\Delta(k+Q_{AF})$ and $\Delta(k)$ are the SC gaps on the nesting electron and hole FSs\cite{Mazin2008,Maier2008,Maier2009}. In this scenario, the newly discovered double spin resonances in NaFe$_{0.985}$Co$_{0.015}$As has been ascribed to anisotropic or orbital dependent SC gaps\cite{Zhang2013,Zhang2014,Zhang2015a,Das2011a}. On the other hand, it has also been suggested that the long range magnetic order can shift the spin resonance to a higher energy at $(\pi, 0)$ from that at the frustrated wave vector $(0, \pi)$ in a detwinned single crystal\cite{Lv2014,Knolle2011}. Hence, the two spin resonances would appear as a double resonance at the same $Q=(\pi, 0)/(0,\pi)$ in a twinned crystal. However, these interpretations could not explain a purely out-of-plane spin excitation enhancement that has been observed at a lower energy ($\approx1.8k_BT_c$) in the SC state in Ba(Fe$_{0.94}$Co$_{0.06}$)$_2$As$_2$, in addition to the commonly observed spin resonance at the energy $\sim4.3k_BT_c$\cite{Steffens2013}. The observation of the out-of-plane spin excitations hints at a connection to the anisotropic spin excitations in the AF parent compound BaFe$_2$As$_2$\cite{Qureshi2012,Wang2014}, where the dispersive spin resonance mode in underdoped Ba(Fe$_{1-x}$Co$_x$)$_2$As$_2$ closely resembles the zone center out-of-plane spin wave mode in the AF state of the parent compound\cite{Christianson2009,Pratt2010,Harriger2009,Kim2013,Lee2013}.

\begin{figure}[b]
\includegraphics[scale=0.5]{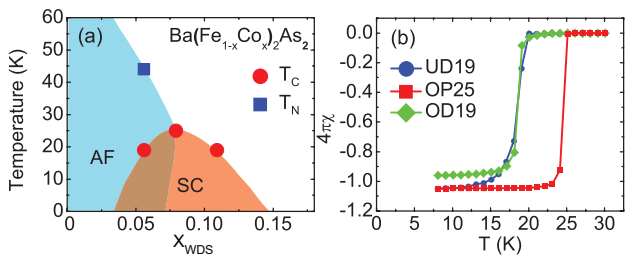}
\caption{ (a) A schematic phase diagram of Ba(Fe$_{1-x}$Co$_x$)$_2$As$_2$ shows $T_N$ and $T_c$ for the compositions $x=$ 0.056, 0.079, and 0.109 samples. (b) Magnetic susceptibility measurements with a 20 Oe field paralleled with the $ab$-plane reveal the $T_c$s at 19, 25, and 19 K for the UD19, OP25, and OD19 samples, respectively.   }
\label{fig1}
\end{figure}

While interpretations of the double spin resonances have been offered in individual cases, a comprehensive explanation across the doping dependence for this peculiar occurrence is lacking\cite{Parshall2009,Matan2009,Lester2010,Li2010}. In this paper, we present a systematic doping dependence study of the Ba(Fe$_{1-x}$Co$_{x}$)$_2$As$_2$ system, where we combine INS and angle resolved photoemission spectroscopy (ARPES) studies on three compounds of the electron doped Ba(Fe$_{1-x}$Co$_x$)$_2$As$_2$ system. We show that the three compounds have comparable SC gaps. Importantly, a three dimensional (3D) spin excitation mode has been observed in the under and optimally doped compounds in the SC state in addition to the commonly observed quasi two dimensional (2D) spin resonance. Moreover, we find that the 3D mode at low energies is monotonically suppressed with doping, while the 2D mode follows the trend of the $T_c$ dome. Hence, we identify the two energy scales of the magnetic excitation enhancements below $T_c$ to be due to the pre-existing magnon mode and the spin exciton mode, respectively. This novel interpretation differs from previous interpretations based on distinct superconducting energy gap scales, and comprehensively resolves the systematic doping-dependence trends.

\section{Experimental details}

The single crystals were grown by the self-flux method\cite{Chen2011}. We determined the compositions using a wavelength-dispersive X-ray spectroscopy (WDS). The $T_c$s were obtained from the onset of the drops in the diamagnetic susceptibilities measured using a standard physical property measurement system (PPMS) from Quantum Design. As shown in Fig. \ref{fig1}, the three compounds that we measured were the underdoped compound $x=0.056, T_N=44, T_c=19$ K (UD19), the optimally doped compound $x=0.079, T_c=25$ K (OP25), and the overdoped compound $x=0.109, T_c=19$ K (OD19). Our neutron scattering experiments were carried out on the HB-3 thermal triple-axis spectrometer at the High Flux Isotope Reactor, Oak Ridge National laboratory. Horizontal collimations of $48^{\prime}-60^{\prime}-80^{\prime}-120^{\prime}$ with a final beam energy of $E_f=14.7$ meV and two pyrolytic graphite (PG) filters were employed.  We coaligned 7.56, 7.39, and 5.41 g of single crystals for each of the compositions, respectively, with a mosaic of $\sim$2$^{\circ}$ full width at half maximum (FWHM) in the $[H, H, L]$ plane in tetragonal notation with the lattice parameters $a=b=3.95$ \AA\ , $c=12.90$ \AA\ for all three compositions optimized at 2 K. The wave vector $Q$ is defined as $Q=[H, K, L]=(2\pi H/a, 2\pi K/b, 2\pi L/c)$ in reciprocal lattice units.  Samples from the same batches were used in the ARPES experiments, which were carried out at beam line 5-4 of the Stanford Synchrotron Radiation Lightsource using a Scienta R4000 analyzer, with a total energy resolution of 5 meV and an angular resolution of 0.3$^{\circ}$. The single crystals were cleaved {\it{in situ}}  below 10 K and measured under ultra high vacuum with a base pressure of better than $3\times10^{-11}$ torr.

\begin{figure}[b]
\includegraphics[scale=0.5]{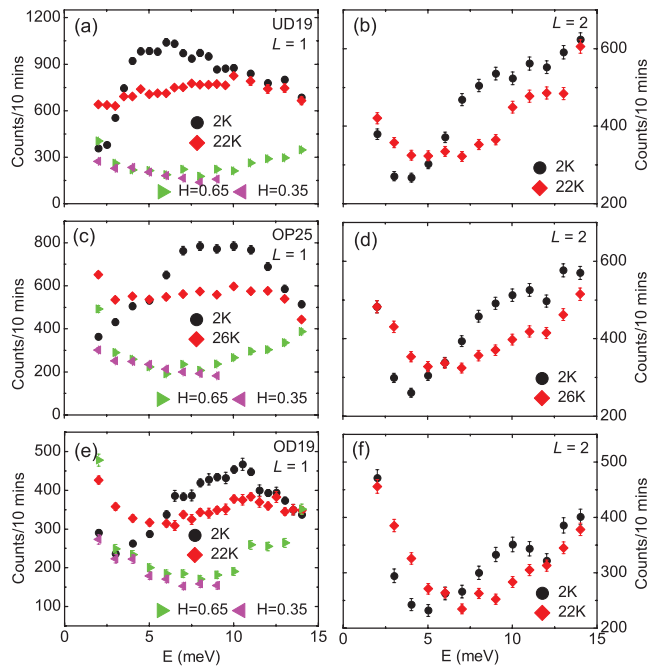}
\caption{ (a) Constant $Q$ scans at $Q=(0.5, 0.5, 1)$, and $Q=(0.65, 0.65, 1)$ and $(0.35, 0.35, 1)$ for background measurements, and (b) at $Q=(0.5, 0.5, 2)$ below and above $T_c$ for the UD19 sample. (c, d) and (e, f) are the identical measurements for the OP25 and OD19 samples.  }
\label{fig2}
\end{figure}

\begin{figure}[t]
\includegraphics[scale=0.5]{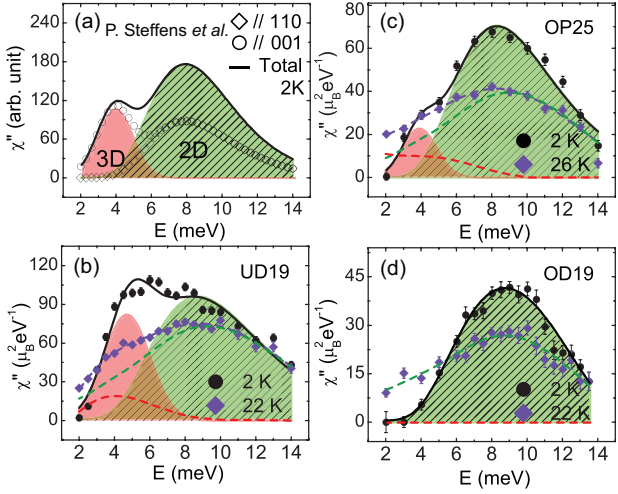}
\caption{ (a) $\chi^{\prime\prime}(\omega, Q)$ measured at $Q=(0.5, 0.5, 1)$ by polarized neutron scattering on Ba(Fe$_{1-x}$Co$_x$)$_2$As$_2$ with $\mathrm{x}=0.06, T_c=24$ K, reproduced from Ref. \onlinecite{Steffens2013}. The red shaded area represents the out-of-plane component, and the stripe green shaded area represents the spin fluctuations presenting in both the in-plane and out-of-plane channels. (b) Normalized $\chi^{\prime\prime}(\omega, Q)$ measured at $Q=(0.5, 0.5, 1)$ on the UD19, (c) OP25, and (d) OD19 compounds with unpolarized INS in the SC state at 2 K (black) and the normal state above $T_c$ (violet). The red and stripe green shaded areas are fits of the spin excitation spectra in the SC state as (a). The black solid lines are sums of the two fitted peaks. The red and olive dashed lines are fits of the 3D and 2D components of the spin excitation spectra in the normal state. The violet dashed lines are sums of the red and olive dashed lines.  }
\label{fig3}
\end{figure}

\begin{figure}[t]
\includegraphics[scale=0.5]{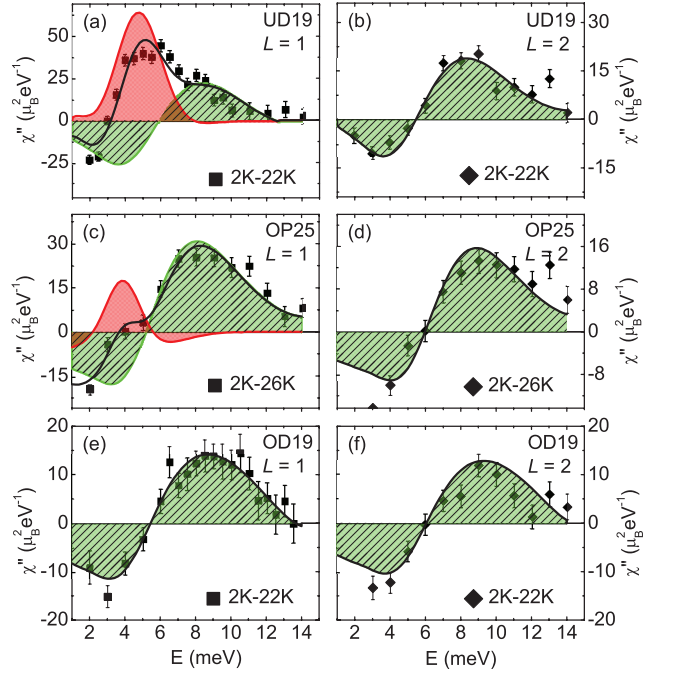}
\caption{ (a) Difference of the normalized total $\chi^{\prime\prime}(Q, \omega)$ between 2 K and 22 K at the wave vectors $Q=(0.5, 0.5, 1)$ and (b) $Q=(0.5, 0.5, 2)$ for the UD19 sample. The red (stripe green) shaded area is the subtraction of the extracted 3D (2D) spin excitations between 2 K and 22 K. The black solid line is a sum of the red and stripe green areas. (c, d) and (e, f) represent the identical analysis for the OP25 and OD19 samples. }
\label{fig4}
\end{figure}

\section{Results}
\subsubsection{Spin excitations}

Figure \ref{fig2} shows constant $Q$ scans of the low energy spin excitations at $Q=(0.5, 0.5, 1)$ and $(0.5, 0.5, 2)$ below and above $T_c$ for all three compounds. The spin excitation spectra reveal clear intensity enhancements below $T_c$, which have been widely accepted as spin resonances. Intriguingly, the spin resonances at $Q=(0.5, 0.5, 1)$ for the UD19 and OD19 samples of the same $T_c$ appear at different energies, deviating from a universal $E_{res}/k_BT_c$ relation. In contrast, the energies of the spin resonances for the three compounds at $Q=(0.5, 0.5, 2)$ are comparable, consistent with the small variation in $T_c$. The deviation of the spin resonance for the underdoped compounds at $L=1$ have been ascribed to a coupling between the magnetic order and superconductivity\cite{Christianson2009,Pratt2010,Steffens2013,Wang2010}. However, the mechanism of the effect of the coupling to the spin resonance is still a puzzle.

Previously, polarized INS studies by P. Steffens {\it et al.} on nearly optimally doped Ba(Fe$_{0.94}$Co$_{0.06}$)$_2$As$_2$ have revealed two components of the spin excitation spectrum with distinct energies, an isotropic in-plane mode and an anisotropic out-of-plane mode\cite{Steffens2013}, as shown in Fig.~\ref{fig3}(a). One can take the difference in intensity between the two polarization channels, thence giving the anisotropic 3D spin excitation mode, while taking the sum of the identical part of the two channels gives the isotropic 2D spin excitation mode thereby separating out the two components. The 3D component can be fitted by a Gaussian peak and the 2D component can be fitted by a Log-normal distribution function. In an unpolarized INS experiment, one measures both the 3D and the 2D spin excitations simultaneously, which in this compound would be seen as a double-peak excitation. 

Figures \ref{fig3}(b)-\ref{fig3}(d) present our unpolarized INS spectra on the UD19, OP25, and OD19 samples below and above $T_c$, respectively. The intensities have been background subtracted, corrected by the Bose factor and normalized by the sample incoherent elastic scattering to give $\chi^{\prime\prime}(Q, \omega)$ in units of $\mu_B^2eV^{-1}$[\onlinecite{Xu2013}]. The suppression of the intensities for the OP25 sample compared with that normalized by phonons in BaFe$_{1.85}$Co$_{0.15}$As$_2$ [Ref. \onlinecite{Inosov2009}] can be ascribed to the contributions of the FeAs flux and Al holders to the incoherent elastic scattering in our normalization. The spin excitation spectrum for the UD19 sample clearly shows two peaks at 2 K, reminiscent of the two components revealed by the polarized INS study. This motivates us to extract the 2D and 3D components of the data both below and above $T_c$ for the three dopings. For below $T_c$, we fit the data with a Gaussian peak and a Log-normal peak, which correspond to the 3D and 2D spin excitations, respectively, for all three dopings. The FWHM of the Gaussian function ($\omega$) and the standard deviation of the Log-normal function ($\sigma$) are two parameters that substantially affect the fittings. From fitting of the polarized INS data in Fig. \ref{fig3}(a), we extract the two parameters, $\omega=2.3\pm0.1$ meV and $\sigma=0.30\pm0.01$, for the $T_c=24$ K sample. For our OP25 sample, which has a very close concentration, we fix $\omega=2.3$ meV and find $\sigma=0.35\pm0.02$. A fitting for the UD19 sample with fixed $\sigma=0.35$ yields $\omega=3.1\pm0.2$ meV. All the other parameters have been released in the fittings. For the OD19 sample, only a Log-normal function is needed to fit the data, yielding $\omega=0.39\pm0.03$. The fitted results as shown in Fig.~\ref{fig3} demonstrate that the 3D excitations become weaker in the OP25 sample, while in the OD19 sample, only the 2D spin excitations remain. For the normal state, knowing that the spin excitations arise purely from the 2D component for the OD19 sample, we have taken the smoothed lineshape of the 2D component from the OD19 sample to fit the 2D contributions for the UD19 and the OP25 samples in the normal state by assuming that the intensities between 10 and 14 meV are purely 2D. The 3D contributions in the normal state can then be obtained by subtracting the fitted 2D component from the total measured $\chi^{\prime\prime}(Q, \omega)$. 

\begin{figure}[t]
\includegraphics[scale=0.45]{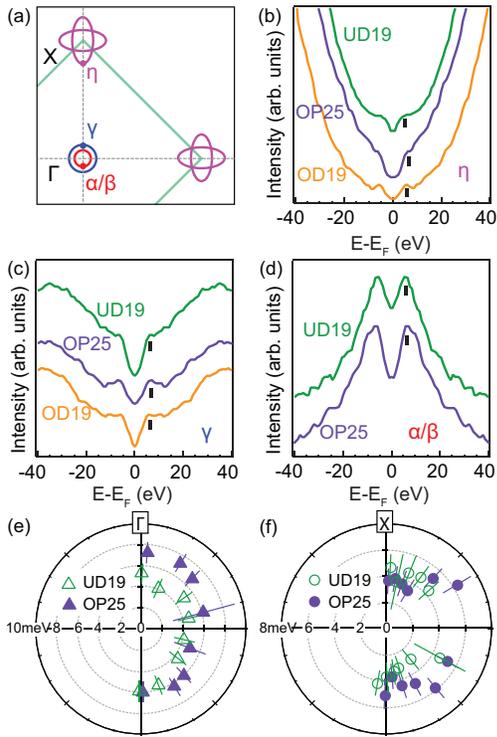}
\caption{(a) A schematic of FSs for Ba(Fe$_{1-x}$Co$_x$)$_2$As$_2$ with the hole FSs ($\alpha, \beta, \gamma$) at $\Gamma$ in the BZ center and the electron FSs ($\eta$) at $\mathrm{X}$ in the BZ corner. The momentum points where the gaps are taken in (b)-(d) are marked on the FSs in (a). Symmetrized EDCs indicating the SC gaps of the UD19 (green), OP25 (purple), and OD19 (orange) samples on the (b) $\eta$ electron band at X, (c) $\gamma$ hole band at $\Gamma$, and (d) inner hole band, $\alpha/\beta$, at $\Gamma$. (b)-(c) were taken with 22 eV photons ($k_z = 0$), while (d) were taken with 30 eV photons ($kz=\pi$) as the inner hole band does not cross the Fermi level at $k_z = 0$. Angular dependence of SC gaps taken at 24 eV for (e) the outer hole band at $\Gamma$, and (f) the electron $\eta$ band at $X$, for the UD19 and OP25 samples.
}
\label{fig5}
\end{figure}

\begin{figure}[b]
\includegraphics[scale=0.55]{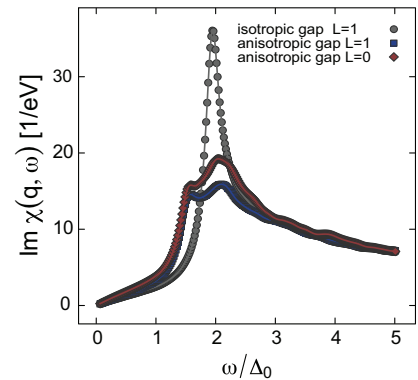}
\caption{ RPA calculation on the spin resonance for the UD19 sample, based on the FSs that have been measured on the sample. The isotropic gap is directly measured by ARPES. For the anisotropic gap, the gap magnitude on the inner electron pocket has been set to be half of that on the outer pocket.}
\label{fig6}
\end{figure}

In order to illustrate the spin excitation response to superconductivity, we present differences of the $\chi^{\prime\prime} (Q, \omega)$ between 2 K and $T>T_c$ at $Q=(0.5, 0.5, 1)$ [Figs. \ref{fig4}a, \ref{fig4}c, \ref{fig4}e] and $Q=(0.5, 0.5, 2)$ [Figs. \ref{fig4}b, \ref{fig4}d, and \ref{fig4}f] for the three samples. The red and the stripe green shaded areas are the differences across $T_c$ for the 3D and the 2D components, respectively, as fitted in Fig.~\ref{fig3}. The spin excitation differences at $L=2$ below and above $T_c$ for all three samples can be described by the corresponding 2D component at $L=1$ with the same lineshape together with an individual intensity scaling factor for each concentration. The disappearance of the 3D mode at $L=2$ in the UD19 and the OP25 samples is consistent with a spin wave mode that evolves to a higher energy and weaker intensities at the Brilliouin zone (BZ) boundary\cite{Pratt2010}. The consistency of the 2D components at $L=1$ and the total spin excitations at $L=2$ across the three dopings further demonstrates that the fitted 3D spin excitations are anisotropic and that the fitted 2D spin excitations are quasi two dimensional in the unpolarized INS spectra. However, the ratio of the maximum intensities at $L=1$ and 2 for the OD19 and the OP25 samples deviate from the evolution of the Fe$^{2+}$ magnetic form factor $0.826/0.756=1.09$, indicating that either the magnetic order couples to superconductivity or that  there is a portion of contributions from the 3D mode at $L=2$ in the two compounds. More interestingly, as revealed in Fig.~\ref{fig4}, the 2D spin excitations are enhanced at the expense of the low energy spin excitation loss, consistent with the opening a spin gap in the SC state. In contrast, there is no such gap for the 3D mode.

\begin{table*}[t]
\caption{Parameters for all the compositions of Ba(Fe$_{1-x}$Co$_x$)$_2$As$_2$ that have been plotted in Fig. \ref{fig7}. The parameters are determined by WDS, INS, and ARPES from our measurements and those in the literature\cite{Qureshi2012,Wang2014,Christianson2009,Pratt2010,Steffens2013,Lumsden2009}. The $2\Delta=\Delta_{\gamma}+\Delta_{\eta}$, because the nesting is between the outer $\gamma$ orbital of the hole FSs and the $\eta$ orbital of the electron FSs. }
\begin{tabular}{cccccccccccc}
\hline \hline
Doping regime & Actual $x$ & Unified $x$& $T_c$ (K) & $T_N$ (K)  & $E_{001}$ (meV)   & $E_{res}$ (meV)&$\Delta_{\alpha/\beta}$ & $\Delta_{\gamma}$&$\Delta_{\eta}$ &$2\Delta$ (meV)   & Ref. \\ \hline
NON    & 0       & 0      & -  & 138  & $13\pm1$    & -    & -& -& -&  - & \onlinecite{Qureshi2012,Wang2014}   \\ 
UD     & 0.04    & 0.042  & 11 & 58   & $7\pm1$     & $4.5\pm0.5$  & & & & &  \onlinecite{Christianson2009}  \\ 
UD     & 0.047   & 0.049  & 17 & 47   & $5.5\pm0.5$ & $5.5\pm0.5$  & & & & & \onlinecite{Pratt2010}    \\ 
UD     & 0.056    & 0.051  & 19 & $44\pm2$ & $4.7\pm0.3$ & $8.3\pm1$ & 5.8 & 5.0 &4.0  & $9\pm1$ &  this work   \\ 
UD     & 0.06    & 0.061  & 24 & -    & $4\pm0.2$   & $8.0\pm1$ & & & &  & \onlinecite{Steffens2013}   \\ 
OP     & 0.079    & 0.067  & 25 & -    & $3.9\pm0.3$ & $8.0\pm1$ &6.5 &6.5 & 4.6&$11.1\pm1$  &   this work   \\ 
OD     & 0.08    & 0.084  & 22 & -    & -       & $8.6\pm0.5$ & & & & &\onlinecite{Lumsden2009}   \\ 
OD     & 0.109   & 0.095  & 19 & -    & -       & $8.5\pm1$  &- &5.6 &4.6 &$10.2\pm1$ &   this work   \\ \hline \hline
\end{tabular}
\label{table:t1}
\end{table*}

\subsubsection{SC gaps and RPA calculation}

As shown above, the energies of the 3D and the 2D spin excitation enhancements in the SC state are distinct. Accordingly, if they both are spin resonance modes, one would expect that the three samples would have SC gaps with distinct or anisotropic energies, especially for the UD19 sample\cite{Das2011a}. We have measured the FSs and SC gaps for all three compositions using ARPES.  The SC gaps are comparable for different bands and compositions, as shown by the symmetrized energy distribution curves (EDCs) in Figs.~\ref{fig5}(b-d), with the gap values shown in Table \ref{table:t1}, which is consistent with previous measurements on the Ba(Fe$_{1-x}$Co$_x$)$_2$As$_2$ system\cite{Terashima2009}. In Figs.~\ref{fig5}(e, f) we present the angular dependence of the SC gaps on the hole FSs around the BZ center, and the electron FS around the BZ corner for the UD19 and the OP25 samples, respectively. The locations in momentum space where these gaps are measured are marked on the FS schematic in Fig. \ref{fig5}(a). The gap magnitude is extracted by fitting the symmetrized EDCs with a symmetrized gap function on top of a Gaussian background for the band at higher binding energy\cite{Norman1998}. The data demonstrate that the SC gaps are largely isotropic within the instrumental resolution, and roughly scale with $T_c$ , consistent with previous reports\cite{Terashima2009}. The SC gaps are consistent with the existence of a spin resonance mode at $8\sim9$ meV, which coincides with the 2D mode, but not the 3D mode which is observed at a much lower energy.

Since the 3D and 2D spin excitation enhancements are most well separated in the UD19 compound, we have carried out random phase approximation (RPA) calculations on the spin resonances for the UD19 sample using a tight-binding model consistent with the electronic structure and isotropic gap measured by ARPES\cite{Maier2008,Maier2009}. There is only one peak in the calculated spin resonance spectrum, as shown in Fig. \ref{fig6}. For comparison, we have also conducted similar calculations by assuming an anisotropic gap on the electron bands at $X$, where the gap magnitude on the inner pocket has been set to be half of that on the outer pocket. The double-peak spin resonance appears as expected at $L=1$\cite{Zhang2013,Zhang2014,Zhang2015a,Das2011a}. However, the result at $L=0$ (this is identical to $L=2$) reveals a qualitatively similar double-peak resonance, which is contrary to the experimentally observed single-peak spin resonance, as shown in Fig. \ref{fig4}. Fine-tuning of the gap anisotropy is required to describe a situation in which the lower resonance is out-of-plane and the higher resonance is in-plane, which makes this explanation unlikely.

\section{Discussion}

\begin{figure}[t]
\includegraphics[scale=0.6]{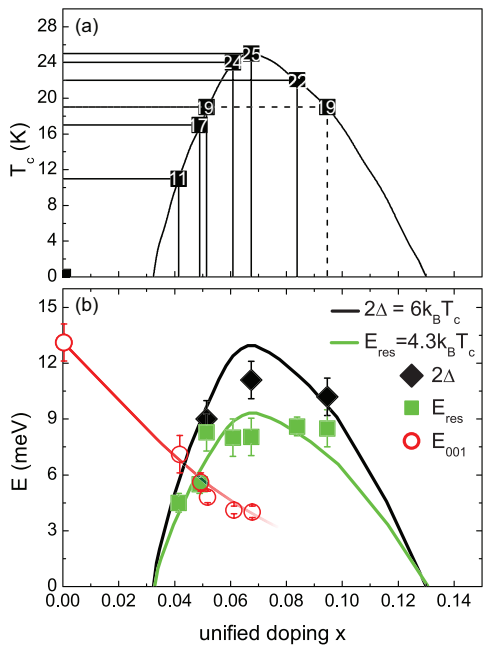}
\caption{ (a) A $T_c-doping$ phase diagram unifies the compositions for all Ba(Fe$_{1-x}$Co$_x$)$_2$As$_2$ samples from this work and the literature\cite{Ni2008,Qureshi2012,Wang2014,Christianson2009,Pratt2010,Steffens2013,Lumsden2009}. The $T_c-doping$ curve is adopted from Ref. \onlinecite{Ni2008}.   (b) A phase diagram showing the energy scales of the SC gap ($2\Delta$), 2D spin resonance ($E_{res}$), and 3D anisotropic spin excitations ($E_{001}$) as a function of the unified doping $\mathrm{x}$ for Ba(Fe$_{1-x}$Co$_x$)$_2$As$_2$. The data for the parent compound, and $T_c=11, 17, 24$ and 22 K compounds are adopted from Ref. \onlinecite{Qureshi2012,Wang2014,Christianson2009,Pratt2010,Steffens2013,Lumsden2009}. The black and green lines show the universal relationships between $2\Delta$, $E_{res}$ and $T_c$ established for the Ba(Fe$_{1-x}$Co$_x$)$_2$As$_2$ system\cite{Yu2009,Park2010,Inosov2011,Terashima2009}. The red line is a guide for the $E_{001}$ mode. }
\label{fig7}
\end{figure}

The lack of a spin gap in Figs.~\ref{fig4} together with the RPA calculations argue that the 3D mode is unlikely to be a real spin resonance mode. Interestingly, previous polarized INS experiments have demonstrated that an out-of-plane spin excitation component at $13\pm1$ meV dominates the anisotropic low energy spin excitation spectrum in the parent material BaFe$_2$As$_2$\cite{Qureshi2012,Wang2014}. This is actually the antiferromagnetic $Q=Q_{AF}$ out-of-plane spin wave mode. The 3D spin excitation modes observed in the under and the optimally doped compounds resemble the out-of-plane spin wave excitations of the parent compound\cite{Qureshi2012,Wang2014}. To determine the origin of the 3D mode, we list the INS studies on the Ba(Fe$_{1-x}$Co$_x$)$_2$As$_2$ system from our own measurements and all those reported in the literature in Table \ref{table:t1}\cite{Christianson2009,Pratt2010,Steffens2013,Lumsden2009,Qureshi2012,Wang2014}. As different reports use different conventions for sample composition, we compile these compounds into a single phase diagram by unifying the doping levels $\mathrm{x}$ by $T_c$ based on a previously established $T_c-doping$ phase diagram\cite{Ni2008}, as illustrated in Fig. \ref{fig7}(a). The SC gaps $2\Delta$ from the sum of the gaps on the hole and electron FSs, the 2D spin resonance energies $E_{res}$, and the energies of the out-of-plane spin wave mode $E_{001}$ as listed in Table \ref{table:t1} are plotted in Fig.~\ref{fig7}(b). The $2\Delta$ and $E_{res}$ follow the scaling relations with the $T_c$ dome that have been established for the Ba(Fe$_{1-x}$Co$_x$)$_2$As$_2$ system\cite{Yu2009,Park2010,Wang2010,Inosov2011,Terashima2009}.

The evolution of the low energy spin excitations in Ba(Fe$_{1-x}$Co$_x$)$_2$As$_2$, on the other hand, is consistent with a monotonic suppression of the out-of-plane spin excitation mode at $13\pm1$ meV in the parent compound\cite{Qureshi2012,Wang2014}, reaching $4\sim5$ meV in the optimally doped compound\cite{Steffens2013}, consistent with the gradual suppression of the long range magnetic order which ultimately becomes short range\cite{Park2010,Lu2014}. This 3D spin excitation mode is well below $2\Delta\approx12$ meV in the optimal doing level; thus the intensities are enhanced due to the suppression of damping in the SC state. The enhanced spin excitations still exhibit three dimensional behavior due to the spatially anisotropic exchange interactions\cite{Harriger2011}.  In this scenario, for a lightly doped SC compound, where the energy of this out-of-plane mode $E_{001}>2\Delta$, the intensities will not be enhanced in the SC state. This is consistent with the observations on the $T_c=11$ K underdoped sample, where a single peak enhancement of the spin excitation is observed below $T_c$, with an energy of $E_{res}=4.5$ meV, as expected for the real spin resonance mode. On the other hand, a second spin excitation mode preexists at 7 meV above $T_c$ and remains unchanged across $T_c$, as the energy of the 3D component is too high ($>2\Delta$) for the removal of damping by superconductivity\cite{Christianson2009}. For a higher doping compound with $T_c=17$ K in the underdoped regime, where $E_{res}\approx E_{001}\approx5.5<2\Delta\approx8.8$ meV, a clear dispersion at the resonance energy has been observed\cite{Pratt2010}. Taking the systematic doping-dependent trends from all compounds together, we conclude that the real spin resonance mode is quasi 2D represented by $E_{res}$, supported by its spin gap and scaling with $T_c$. On the other hand, the dispersive spin resonance observed in the underdoped regime is almost certainly associated with the spin excitations of the magnetic order. The intensity of this mode is enhanced below $T_c$ due to the suppression of the damping. In other words, the anisotropic spin excitation enhancement in the SC state observed in INS spectrum does not arise from the spin exciton mode, and the real spin resonance mode is quasi two dimensional for the three doping levels. We note that there is an $L$ modulation in the intensities for the 2D spin resonance in the underdoped regime. The interpretation also applies to the double spin resonances observed in the electron-doped cuprate, Nd$_{2-x}$Ce$_x$CuO$_{4+\delta}$ with $x\approx0.155$ and $T_c=25$ K\cite{Yu2010}. 

\section{Conclusion}

In summary, we have carried out a combined study of INS, ARPES, and RPA calculations on the two spin excitation enhancement modes in the SC state in Ba(Fe$_{1-x}$Co$_x$)$_2$As$_2$. The ARPES measurements combined with the RPA calculations reveal that the spin singlet-triplet excitations can only account for the spin excitation enhancement at $E\approx4.3k_BT_c$. The analysis of our measurements taken together with the comprehensive measurements in literature strongly suggest that the lower energy spin excitation mode corresponds to the anisotropic spin excitations of the ordered magnetic moments. The characteristic energy monotonically decreases from BaFe$_2$As$_2$ until the magnetic order disappears in the overdoped regime. For the compositions where the spin excitation mode energy falls below $2\Delta$, the intensities of this mode is enhanced in the SC state due to the suppression of damping, and this spin wave excitation resembles  a spin resonance mode. This novel interpretation of the two energy scales of the magnetic excitation enhancements in the superconducting state differs from previous interpretations ascribing them to anisotropic energy gaps, and comprehensively explains the systematic trend across the doping-dependent phase diagram.

\section{Acknowledgments}

This work was supported by the Director, Office of Science, Office of Basic Energy Sciences, Materials Sciences and Engineering Division, of the U.S. Department of Energy under Contract No. DE-AC02-05-CH11231 with the Quantum Materials Program (KC2202) and the Office of Basic Energy Sciences U.S. DOE Grant No. DE-AC03-76SF008. The RPA calculations were conducted at the Center for Nanophase Materials Sciences, which is a DOE Office of Science User Facility. The work at Rice was supported by the U.S. NSF-DMR-1308603 and DMR-1362219 (P.D.), and by
Robert A. Welch Foundation grant no. C-1839 (P.D.). The neutron scattering experiments at Oak Ridge National Laboratory's High-Flux Isotope Reactor were sponsored by the Scientific User Facilities Division, Office of Basic Energy Sciences, U.S. Department of Energy. The ARPES work was performed at the Stanford Synchrotron Radiation Lightsource, which is operated by the Office of Basic Energy Science, U.S. Department of Energy. 
\bibliography{mengbib}

\end{document}